%% file: BPH-16-002_temp.tex
\pdfoutput=1

\documentclass[11pt,twoside,a4paper,cmspaper,final,collab]{cms-tdr}
\def\svnVersion{461751P}\def\cmsCernNoTag{CERN-EP-2017-287}\def\cmsCernDate{\today}\def\cmsMessage{Published in Physical Review Letters as \href{http://dx.doi.org/10.1103/PhysRevLett.120.202005}{\doi{10.1103/PhysRevLett.120.202005.}}}

\begin{document}\cmsNoteHeader{BPH-16-002}

\hyphenation{had-ron-i-za-tion}
\hyphenation{cal-or-i-me-ter}
\hyphenation{de-vices}

\RCS$Revision: 446858 $
\RCS$HeadURL: svn+ssh://svn.cern.ch/reps/tdr2/papers/BPH-16-002/trunk/BPH-16-002.tex $
\RCS$Id: BPH-16-002.tex 446858 2018-02-19 22:09:13Z alverson $
\newlength\cmsFigWidth
\ifthenelse{\boolean{cms@external}}{\setlength\cmsFigWidth{0.85\columnwidth}}{\setlength\cmsFigWidth{0.4\textwidth}}
\ifthenelse{\boolean{cms@external}}{\providecommand{\cmsLeft}{top\xspace}}{\providecommand{\cmsLeft}{left\xspace}}
\ifthenelse{\boolean{cms@external}}{\providecommand{\cmsRight}{bottom\xspace}}{\providecommand{\cmsRight}{right\xspace}}
\ifthenelse{\boolean{cms@external}}
{\providecommand{\suppMaterial}{the supplemental material
[URL will be inserted by publisher]}}
{\providecommand{\suppMaterial}{Appendix~\ref{app:suppMat}}}

\ifthenelse{\boolean{cms@external}}{\providecommand{\CL}{C.L.\xspace}}{\providecommand{\CL}{CL\xspace}}
\providecommand{\PX}{\ensuremath{\cmsSymbolFace{X}}\xspace}
\providecommand{\PBstzs}{\ensuremath{\cmsSymbolFace{B}^{*0}_\cmsSymbolFace{s}}\xspace}
\providecommand{\PBstz}{\ensuremath{\cmsSymbolFace{B}^{*0}}\xspace}
\providecommand{\PBstornotz}{\ensuremath{\cmsSymbolFace{B}^{(*)0}}\xspace}
\providecommand{\PBc}{\ensuremath{\cmsSymbolFace{B}_\cmsSymbolFace{c}}\xspace}
\newcommand{\mBs}       {\ensuremath{m_{\PBzs}}\xspace}
\newcommand{\mBsst}     {\ensuremath{m_{\PBstzs}}\xspace}
\newcommand{\bsudX}     {\ensuremath{\PQb\PAQs\PQu\PAQd}\xspace}
\newcommand{\Xm}      {\ensuremath{\PX(5568)}\xspace}
\newcommand{\Xmpm}      {\ensuremath{\Xm^{\pm}}\xspace}

\cmsNoteHeader{BPH-16-002}
\title{Search for the \texorpdfstring{$\PX(5568)$}{X(5568)} state decaying into
\texorpdfstring{$\PBzs\Pgppm$ \\}{B0s pi+-} in proton-proton collisions at
\texorpdfstring{$\sqrt{s} = 8\TeV$}{sqrt(s) = 8 TeV}}

\date{\today}

\abstract{
A search for resonance-like structures in the $\PBzs\Pgppm$ invariant mass spectrum is performed using
proton-proton collision data collected by the CMS experiment at the LHC
at $\sqrt{s} = 8\TeV$, corresponding to an integrated luminosity of 19.7\fbinv.
The $\PBzs$
mesons are reconstructed in the decay chain
$\PBzs \to \PJGy\,\phi$, with
$\PJGy \to \PGm^+\PGm^-$ and $\phi\to\PKp\PKm$.
The $\PBzs\Pgppm$ invariant mass distribution
shows no statistically significant peaks for different selection requirements
on the reconstructed $\PBzs$ and $\Pgppm$ candidates.
Upper limits are set on the relative production rates of the $\PX(5568)$ and
$\PBzs$ states times the branching fraction of the
decay $\PX(5568)^{\pm} \to \PBzs \Pgppm $.
In addition, upper limits are obtained as a function of the mass
and the natural width of possible exotic states decaying into
$\PBzs\Pgppm$.
}

\hypersetup{%
pdfauthor={CMS Collaboration},%
pdftitle={Search for the X(5568) state decaying into B0s pi+- in proton-proton collisions at sqrt(s) = 8 TeV},%
pdfsubject={CMS},%
pdfkeywords={CMS, physics, b-mesons, tetraquark, exotic hadrons, exotic resonance, exotic mesons, b-physics}}

\maketitle

The evidence presented by the D0 Collaboration of a new state decaying
to $\PBzs\Pgppm$~\cite{D0Obs} initiated considerable interest within
the exotic hadron community
(discussed, e.g., in Refs.~\cite{Burns:2016gvy, Guo:2016nhb}
and references therein) and
triggered a similar search by the LHCb Collaboration~\cite{LHCbpap}.
The D0 experiment reported an unexpected, narrow structure, named $X(5568)$,
in the $\PBzs\Pgppm$ invariant mass distribution
and interpreted it as a hadron composed of four quarks of different
flavors~($\bsudX$; inclusion of charge-conjugate modes
is implied throughout this Letter).
The measured mass and natural width of this state are
$5567.8\pm 2.9\stat {}^{+0.9}_{-1.9}\syst \MeV$
and $21.9\pm 6.4\stat {}^{+5.0}_{-2.5}\syst \MeV$, respectively~\cite{D0Obs}.
Possible quantum numbers for the state are $J^P = 0^+$,
if the $\PBzs\Pgppm$ is produced in an $S$-wave,
or $J^P = 1^+$, if the decay proceeds via the chain
$\Xmpm\to\PBstzs\Pgppm$, $\PBstzs\to\PBzs\gamma$ and the photon is not reconstructed.
In the latter case, the mass of the new state would be shifted
by $m_{\PBstzs}-m_{\PBzs}$ with respect to the measured $\Xm$ mass, where $m_{\PBzs}$
and $m_{\PBstzs}$ are the nominal $\PBzs$ and $\PBstzs$ masses~\cite{PDG}.

The LHCb Collaboration searched for the $\Xm$ state and reported
a negative result~\cite{LHCbpap}.
Further independent searches are needed either to confirm
the $\Xm$ state or to set stronger limits on its production.
In particular, the CMS detector can probe a central kinematic region of
$\PBzs$ candidates similar to that of D0, complementing the LHCb
search in the forward region.
Recently, the CDF and ATLAS Collaborations reported independently
negative search results for the $\Xm$~\cite{Aaltonen:2017voc,Aaboud:2018hgx},
while the D0 Collaboration presented additional evidence for the $\Xm$
by adding $\PBzs$ mesons reconstructed in semileptonic decays~\cite{Abazov:2017poh}.

This Letter presents a search for the $\Xm$ state performed by
the CMS Collaboration at the LHC. The data sample corresponds to
19.7\fbinv of proton-proton ($\Pp\Pp$) collisions
at $\sqrt{s}= 8\TeV$ collected in 2012.
The $\PBzs\Pgppm$ candidates are reconstructed through the decay
$\PBzs\to\PJGy\,\phi$, with $\PJGy\to\PGmp\PGmm$ and $\phi\to\PKp\PKm$.
The relative production rate of $\Xm$, with
respect to $\PBzs$, times the branching fraction
of the $\Xmpm\to\PBzs\Pgppm$ decay is calculated using the relation
\ifthenelse{\boolean{cms@external}}
{
\begin{linenomath*}
\begin{equation}\label{eqnrho}
\begin{aligned}
\rho_{\PX} & \equiv \frac{\sigma(\Pp\Pp \to \PX + \text{anything}) \,
{\mathcal{B}}(\PX\to\PBzs\Pgppm)}{\sigma(\Pp\Pp \to \PBzs + \text{anything})} \\
& = \frac{N_{\PX}}{\epsilon_{\text{rel}} N_{\PBzs}}, \end{aligned} \end{equation}
\end{linenomath*}
}
{
\begin{linenomath*}
\begin{equation}\label{eqnrho}
\rho_{\PX} \equiv \frac{\sigma(\Pp\Pp \to \PX + \text{anything})
\, {\mathcal{B}}(\PX\to\PBzs\Pgppm)
}
{\sigma(\Pp\Pp \to \PBzs + \text{anything})
}
= \frac{N_{\PX}}{\epsilon_{\text{rel}} N_{\PBzs}},
\end{equation}
\end{linenomath*}
}
where $\PX=\Xmpm$, $N_{\PX}$~($N_{\PBzs}$) is the number of $\Xm~(\PBzs)$ signal
candidates reconstructed in data and
$\epsilon_{\text{rel}} = \epsilon_{\PX}/\epsilon_{\PBzs}$
is the relative efficiency.
The D0 Collaboration measured $\rho_{\PX}=(8.6\pm2.4)\%$
and $(8.2\pm3.1)\%$ for $\pt(\PBzs)>10$ and 15\GeV~\cite{D0Obs}.

The central feature of the CMS apparatus is a superconducting solenoid
of 6\unit{m} internal diameter, providing a magnetic field of 3.8\unit{T}.
Within the solenoid volume are a silicon pixel and strip tracker, a lead
tungstate crystal electromagnetic calorimeter, and a brass and
scintillator hadron calorimeter, each composed of a barrel and two
endcap sections. Muons are detected
in the pseudorapidity range  $\abs{\eta}<2.4$
in gas-ionization chambers embedded in
the steel flux-return yoke outside the solenoid.
The main subdetectors used for the present analysis are
the silicon tracker and the muon detection system.
The silicon tracker measures charged particles within the
range $\abs{\eta} < 2.5$.
For nonisolated particles with transverse momentum $1 < \pt < 10\GeV$
and $\abs{\eta} < 1.4$,
the track resolutions are typically 1.5\% in \pt and 25--90\,(45--150)\mum
in the transverse (longitudinal) impact parameter~\cite{TRK-11-001}.
Matching muons to tracks measured in the silicon tracker
results in a relative $\pt$ resolution for muons
with $\pt < 10\GeV$ of 0.8--3.0\% depending on
$\abs{\eta}$~\cite{Chatrchyan:2012xi}.
A more detailed description of the CMS detector, together with a definition
of the coordinate system used and the relevant kinematic variables,
can be found in Ref.~\cite{Chatrchyan:2008zzk}.

Events of interest are selected using a two-tiered trigger
system~\cite{Khachatryan:2016bia}. The first level (L1), composed of custom
hardware processors, uses information from the calorimeters and muon detectors.
The second level, known as the high-level trigger
(HLT), consists of a farm of processors running a version of the full event
reconstruction software optimized for fast processing.
This analysis uses events collected with HLT algorithms requiring
two muons that are consistent with originating from a $\PJGy$ meson
decaying at a significant distance from the luminous region.

The reconstruction of the $\PBzs$ candidates closely follows the
procedure described in Ref.~\cite{BsCPVCMS}, where the CP-violating
phase $\phi_s$ was measured using
the same decay chain, $\PBzs\to\PJGy\,\phi$, with $\PJGy\to\PGmp\PGmm$ and $\phi\to\PKp\PKm$,
reconstructed from the same data set and triggered by the same
L1 and HLT algorithms.

The reconstruction requires two muons of opposite charge
that must match those that triggered the event readout.
The offline muon selection is more restrictive
than the trigger requirements and includes:
$\pt(\PGmpm)>4\GeV$, $|\eta(\PGmpm)|<2.2$, $\pt(\PGmp\PGmm)>7\GeV$,
soft muon identification~\cite{Chatrchyan:2012xi},
the dimuon vertex $\chi^2$ fit probability $P_{\mathrm{vtx}}(\PGmp\PGmm) > 10\%$,
and the dimuon mass within the range 3.04 to $3.15\GeV$.
The angle $\alpha_{\mathrm{T}}(\PGmp\PGmm)$ in the transverse plane
between $\vec{p}_{\mathrm{T}}(\PGmp\PGmm)$ and the vector $\vec{D}_{xy}(\PGmp\PGmm)$ from
the beam axis to the dimuon vertex is required to
satisfy $\cos{\alpha_{\mathrm{T}}(\PGmp\PGmm)} > 0.9$.
The dimuon vertex transverse displacement divided by its uncertainty,
$D_{xy}(\PGmp\PGmm)/ \sigma_{D_{xy}(\PGmp\PGmm)}$, must be greater than 3.

The two muons are combined with two other oppositely charged
tracks in the event, neither identified as a muon and assumed to
be kaons, which must each have
$\pt(\PKpm)>0.7\GeV$ and satisfy high-purity
track~\cite{TRK-11-001} requirements.
The invariant mass of the kaon pair candidate, $M(\PKp\PKm)$, is
required to be within ${\pm}10\MeV$ of the
known $\phi(1020)$ meson mass~\cite{PDG}.

The $\PBzs$ candidates are obtained using a kinematic vertex fit
to the two muon and two kaon tracks,
with the dimuon candidate mass constrained to
the nominal $\PJGy$ meson mass~\cite{PDG}
(the mass of the $\PKp\PKm$ candidate is not constrained because the
width of the $\phi(1020)$ resonance exceeds the mass resolution).
Additional requirements imposed on the $\PBzs$ candidates include
$\pt(\PBzs)>10\GeV$, $P_{\mathrm{vtx}}(\PBzs)>1\%$,
$D_{xy}(\PBzs)/\sigma_{D_{xy}(\PBzs)}>3$,
and $\cos{\alpha_{\mathrm{T}}(\PBzs)}>0.99$,
where $D_{xy}(\PBzs)$ and $\alpha_{\mathrm{T}}(\PBzs)$ are analogous
to the corresponding dimuon variables and are measured
with respect to the primary interaction vertex (PV).
The events contain multiple $\Pp\Pp$ collisions from the same or nearby
bunch crossings (pileup), with an average of 16 collisions per event.
The PV is chosen as the one with the smallest angle between the
vector from the collision point to the $\PBzs$ candidate decay vertex
and the $\PBzs$ candidate momentum.

\begin{figure}[bt]
\begin{center}
\includegraphics[width=0.48\textwidth] {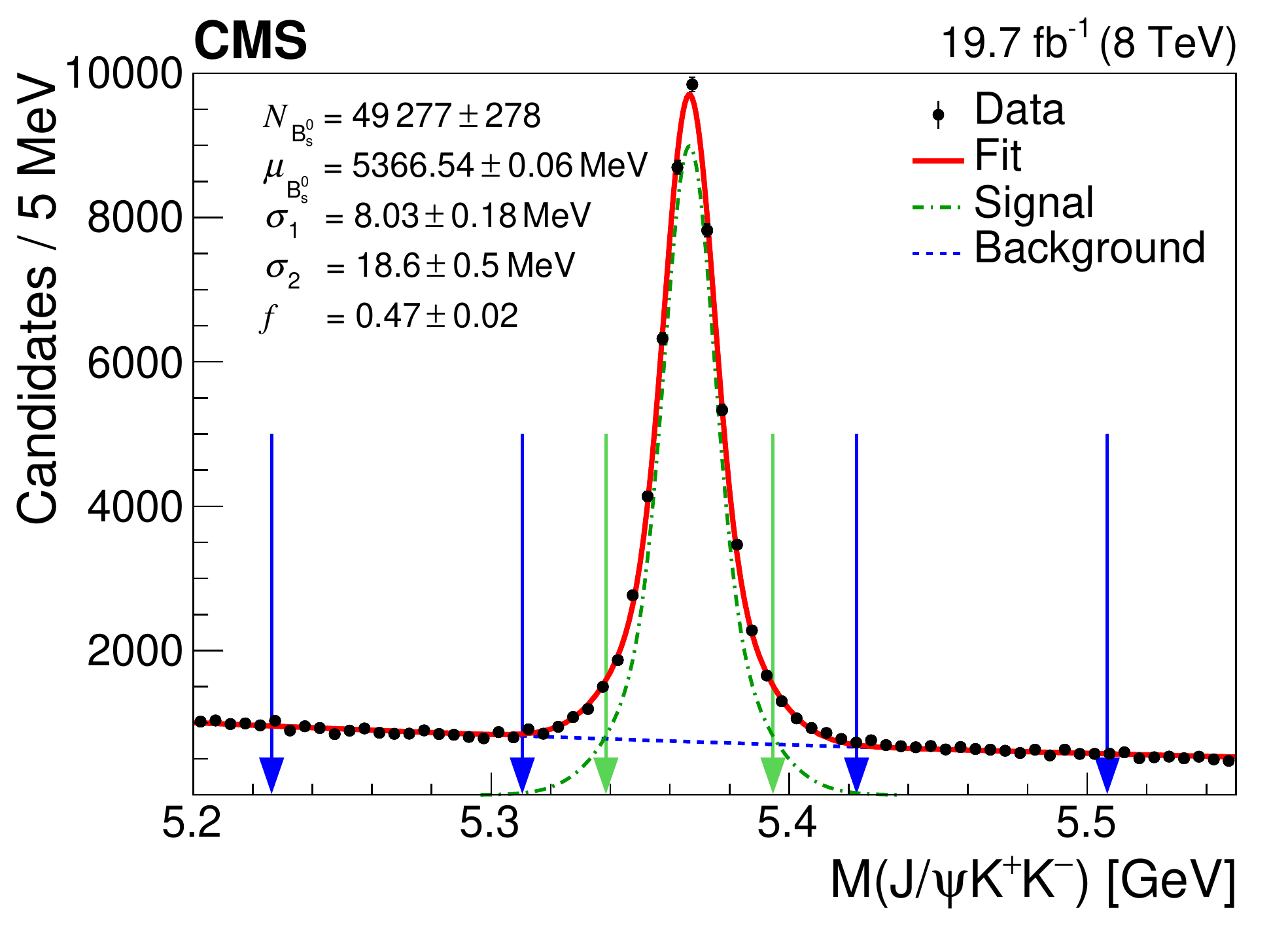}
\end{center}
\caption{Invariant mass distribution of the $\PBzs$ candidates with the fit
result superimposed.
The outermost pairs of dark vertical arrows define the lower and upper $\PBzs$
sidebands, while the innermost light vertical arrows delimit the signal region.
\label{figBs}}
\end{figure}

An extended unbinned maximum-likelihood fit to the $ \PJGy \PKp \PKm $
invariant mass, $M(\PJGy\PKp\PKm)$, distribution yields $49\,277\pm278$
$\PBzs$ signal candidates, where the signal and background
components are modeled by a double-Gaussian and an exponential
function, respectively, as shown in Fig.~\ref{figBs}.
In the fit, the common mean ($\mu_{\PBzs}$), the fraction of the
second Gaussian function ($f$), and the widths ($\sigma_{1,2}$) of
the two signal Gaussian functions (given in Fig.~\ref{figBs}), as well as
the parameter of the exponential function, are left free.
Signal and lower and upper sideband mass regions are defined, respectively,
by the intervals $[-2\sigma_{\PBzs},\,+2\sigma_{\PBzs}]$, $[-10\sigma_{\PBzs},\,-4\sigma_{\PBzs}]$, and
$[\,4\sigma_{\PBzs},\,10\sigma_{\PBzs}]$ around $\mu_{\PBzs}$,
as indicated in Fig.~\ref{figBs}.
Here, $\sigma_{\PBzs}\simeq14\MeV$ represents
the standard deviation of the double-Gaussian function.
In the signal region, the signal purity is about 85\%
and the number of multiple $\PBzs$ candidates in a single event is
negligible.

The pion candidate from the $\Xmpm\to\PBzs\Pgppm$ decay is required to
be a track used in the PV fit, with $\pt(\Pgppm) > 0.5\GeV$, and satisfy
track quality requirements~\cite{TRK-11-001}.
The average number of $\PBzs\Pgppm$ candidates per event in the
$\PBzs$ signal region is 1.8.
Constraints on the angle between the momenta of the $\PBzs$ and $\Pgppm$
candidates are not imposed in this analysis, because such requirements
sculpt the $\PBzs\Pgppm$ invariant mass distribution in a nontrivial way,
as discussed in \suppMaterial.
Motivated by the low momentum of the $\Pgppm$ from the signal decay,
together with the strong correlation between $M(\PJGy\PKp\PKm \Pgppm)$
and $M(\PJGy\PKp\PKm)$, the invariant mass of a $\PBzs\Pgppm$ candidate is defined as
$M^{\Delta}(\PBzs\Pgppm) =  M(\PJGy\PKp\PKm \Pgppm) - M(\PJGy\PKp\PKm) + m_{\PBzs}$.
This improves the $\PBzs\Pgppm$ mass resolution by a factor of 5,
as found in Monte Carlo (MC) simulations.

Simulated MC events are produced with \textsc{Pythia}
v6.424~\cite{Sjostrand:2006za}.
The $\Xm$ is simulated as a spin-0 particle of mass and width equal to
$m_{\PX} \equiv 5567.8\MeV$ and $\Gamma_{\PX} \equiv 21.9\MeV$, and is
forced to decay to a $\PBzs$ meson and a $\Pgppm$ using the phase-space model
in {\sc EvtGen} 1.3.0~\cite{Lange2001152}.
For the $\PBzs$ signal generation, {\sc EvtGen} simulates
the $\PBzs\to\PJGy\,\phi$, $\PJGy\to\PGmp\PGmm$, and $\phi\to\PKp\PKm$ decays, including the
effects from mixing and CP violation.
Final-state photon radiation is included in \textsc{EvtGen} using
\textsc{PHOTOS}~\cite{BARBERIO1991115,BARBERIO1994291}.
The events are then passed through a detailed \textsc{Geant4}-based
simulation~\cite{Agostinelli2003250} of the CMS detector with the
same trigger and reconstruction algorithms used on data.
The simulation includes pileup effects at the same rate as observed in data.

\begin{figure*}[btp]
\begin{center}
\includegraphics[width=0.48\textwidth]{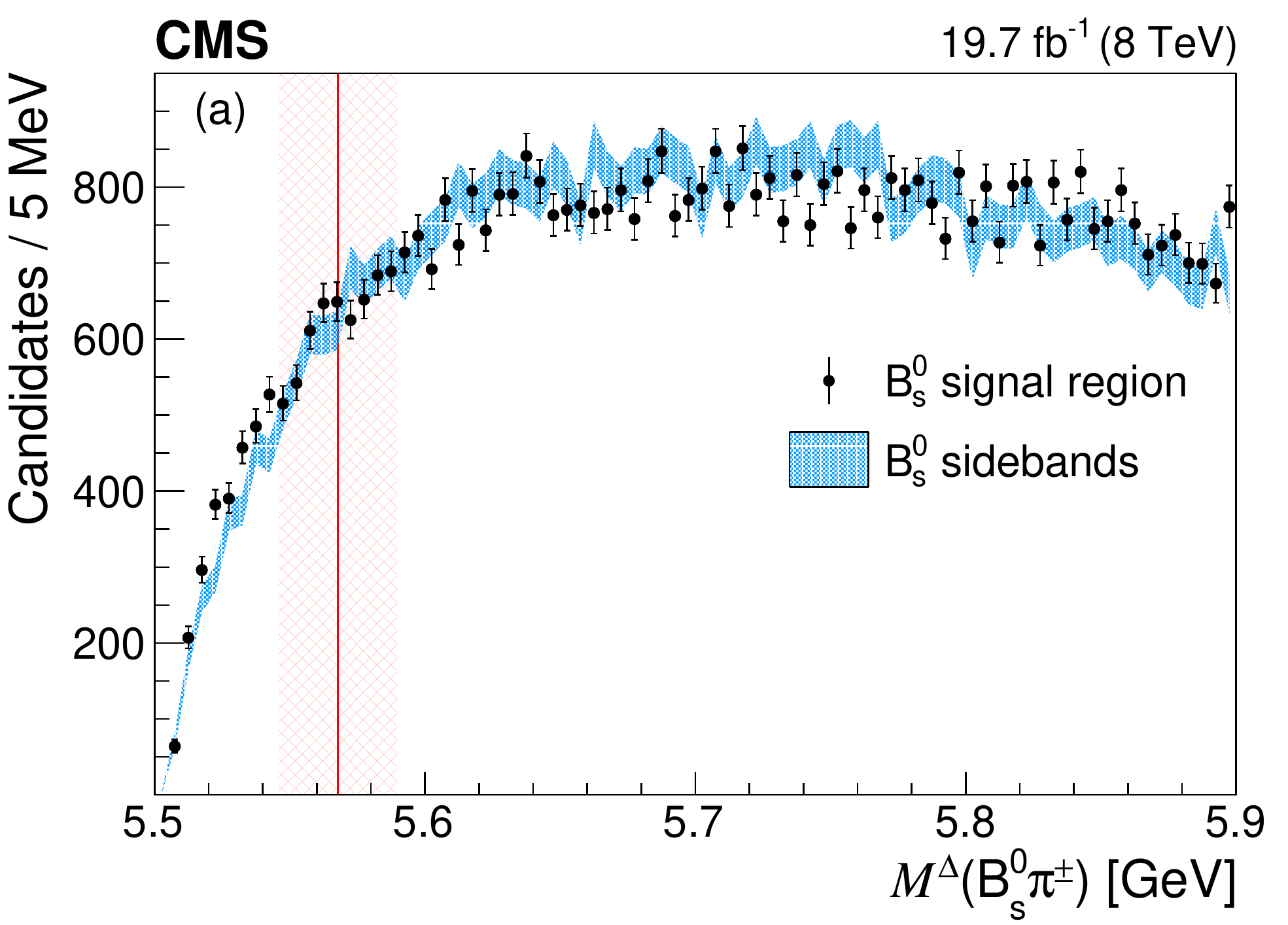}
\includegraphics[width=0.48\textwidth]{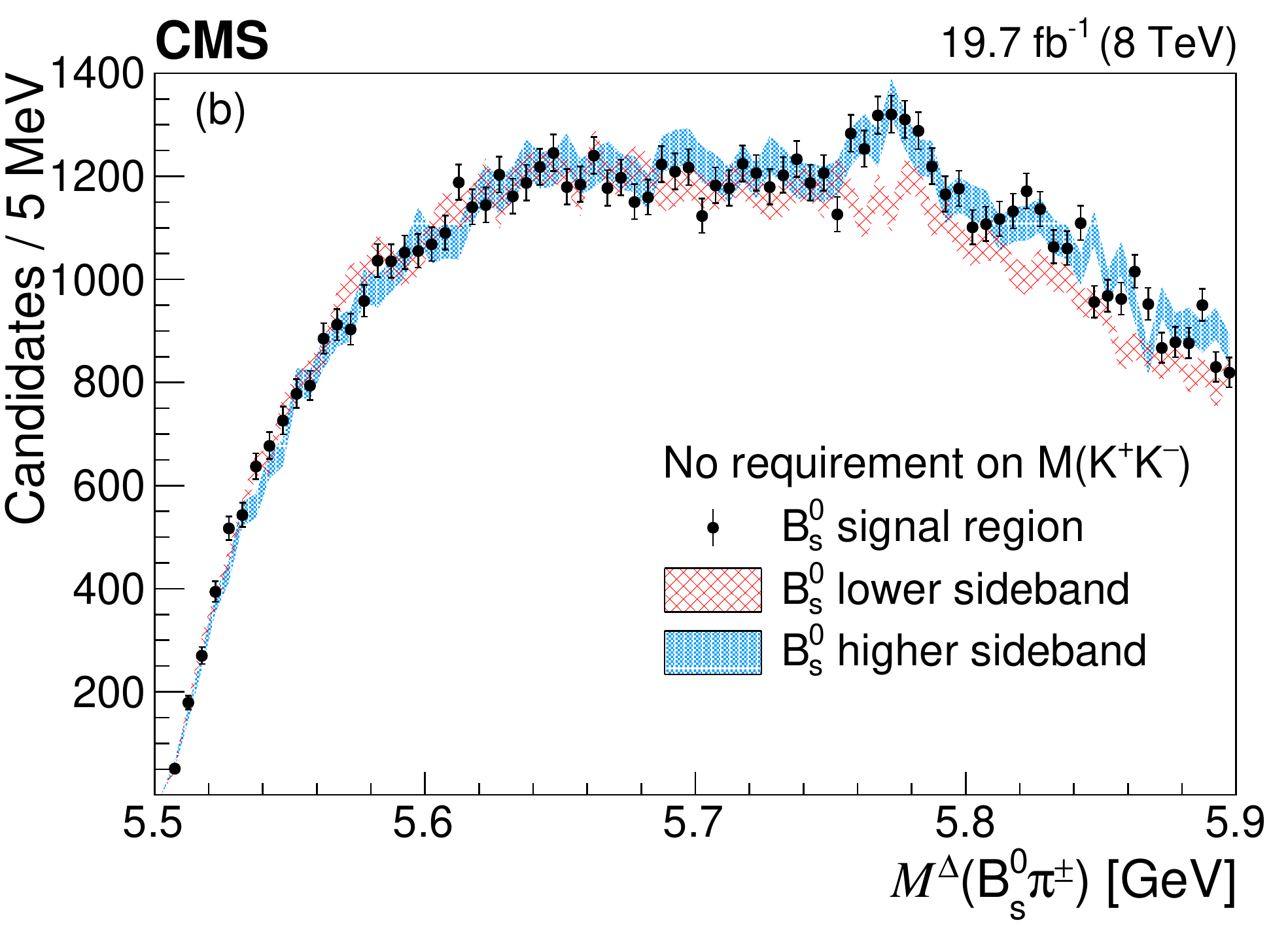}
\end{center}
\caption{
(a) The $M^{\Delta}(\PBzs\Pgppm)$ distribution for events in the $\PBzs$
signal (points) and sideband regions (bands).
The latter is normalized to the former.
The vertical band indicates the region $m_{\PX}\pm\Gamma_{\PX}$.
(b) The $M^{\Delta}(\PBzs\Pgppm)$ distribution for events in the $\PBzs$ signal (points)
and lower and higher sideband regions (bands), when the requirement
on $ M(\PKp\PKm) $ is removed (see text for additional requirements).
The three distributions are normalized from the mass threshold
up to $5.74\GeV$.
The excess observed for events in the $ \PBzs $ signal and higher sideband
regions is due to $\PB^{(*)}_{1,2}{}^+\to\PBstz\Pgpp$ decays.
\label{BspiSigvsBkg}}
\end{figure*}
The $M^{\Delta}(\PBzs\Pgppm)$ distributions obtained from events in  the $\PBzs$ signal
and sideband regions are compared after normalization in
Fig.~\ref{BspiSigvsBkg}(a), showing no significant differences
from threshold near $5.5\GeV$ up to $5.9\GeV$.
In particular, no excess is visible near $5568\MeV$.

To verify the reconstruction procedure, the
requirement on $M(\PKp\PKm)$ is removed.
This allows the $\PBz\to\PJGy\PKp\Pgpm$ decay to contribute to the
resulting $M(\PJGy\PKp\PKm)$ distribution, but only in the $\PBzs$ signal
and the higher sideband regions, as verified by simulation.
Additional requirements are imposed to reduce
the level of background:
$\pt(\PBzs)>25\GeV$, $\pt(\Pgppm)>1\GeV$, and $\pt(\PKpm)>1\GeV$.
Figure~\ref{BspiSigvsBkg}(b) shows the
resulting $M^{\Delta}(\PBzs\Pgppm)$ distributions
for events in the lower and higher sideband and signal regions.
Only the latter two distributions have a clear excess around $5.75{-}5.84\GeV$.
This excess is consistent with the decays $\PB_{1}(5721)^+\to\PBstz\Pgpp$,
$\PB^{*}_{2}(5747)^+\to\PBstz\Pgpp$, and $\PB^{*}_{2}(5747)^+\to\PBz\Pgpp$,
where the decay $\PBz\to\PJGy\PKp\Pgpm$ is misreconstructed as $\PBzs\to\PJGy\PKp\PKm$
(the photon from the $\PBstz\to\PBz\gamma$ decay is not reconstructed).
The peaks in the $ M^{\Delta}(\PBzs\Pgppm) $ distribution corresponding to the
decays $\PB^{(*)}_{1,2}{}^+\to\PBstz\Pgpp$ are shifted by $m_{\PBzs}-m_{\PBstornotz}$
with respect to the nominal masses of the $\PB^{(*)}_{1,2}{}^+$ states~\cite{PDG}.

\begin{figure*}[tb]
\begin{center}
\includegraphics[width=0.48\textwidth]{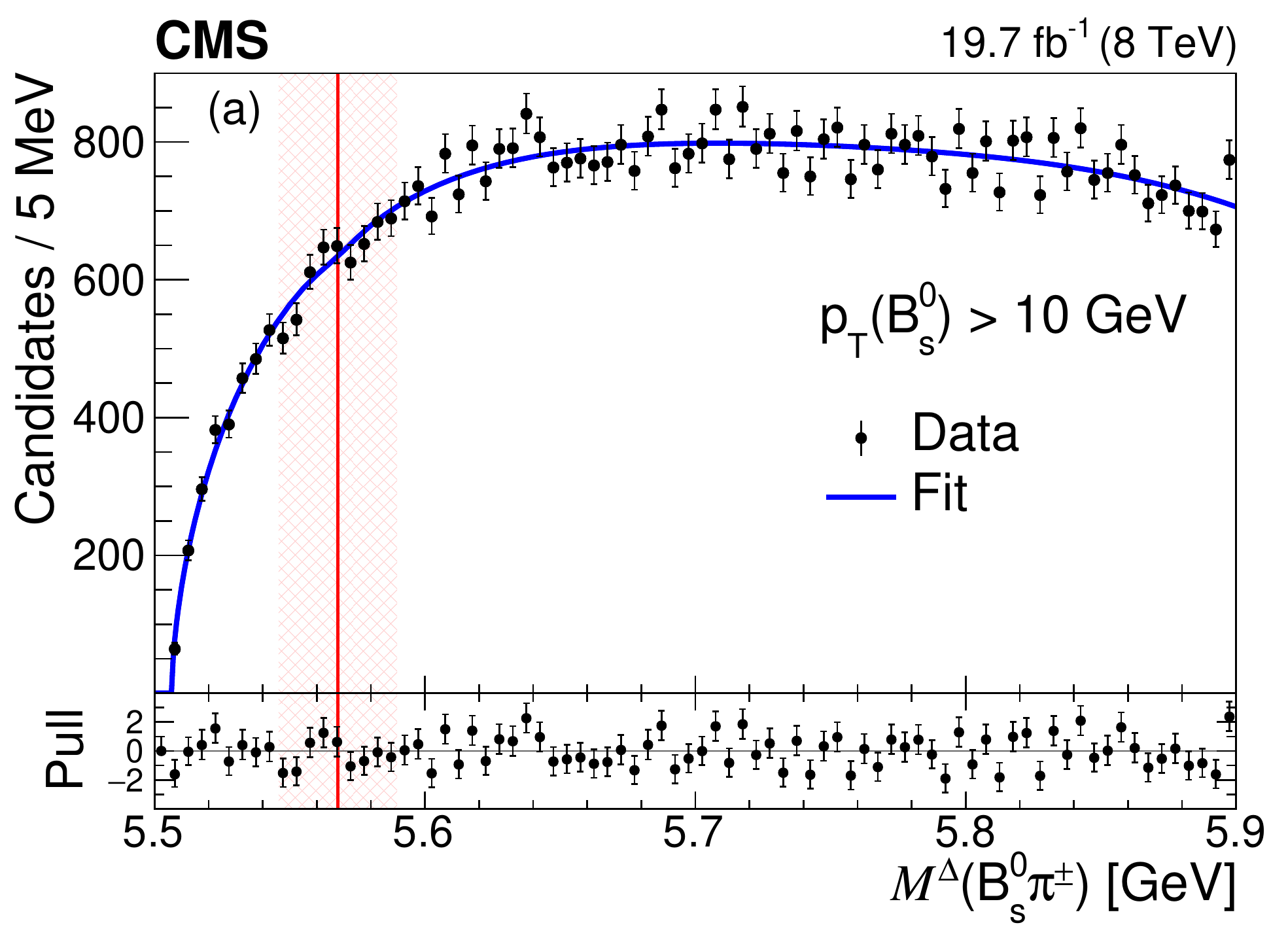}
\includegraphics[width=0.48\textwidth]{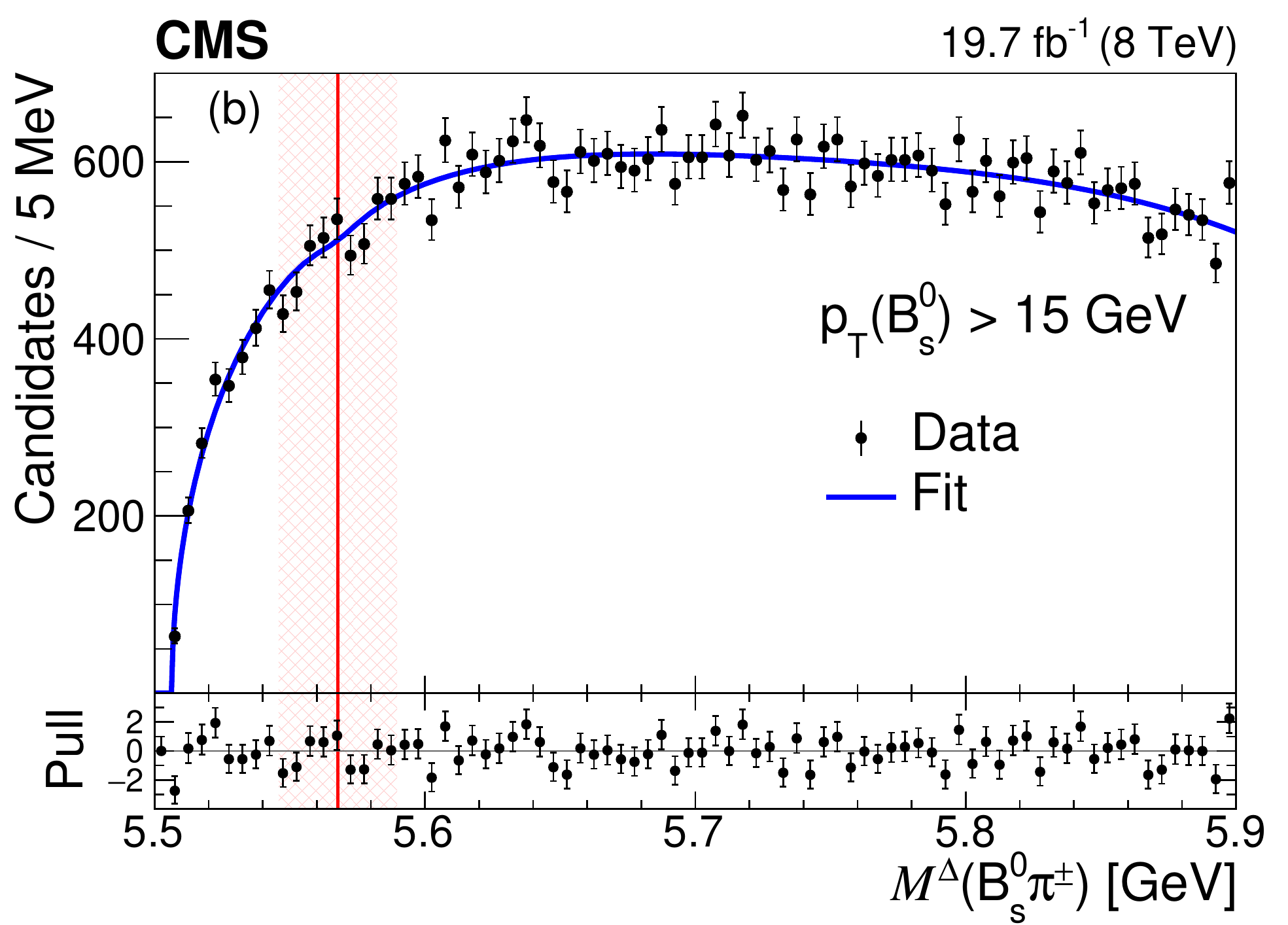}
\end{center}
\caption{
The $M^{\Delta}(\PBzs\Pgppm)$ distribution for events in the $ \PBzs $ signal region with the
result of the fit superimposed for the baseline selection with
$\pt(\PBzs)>10\GeV$ (a) and $\pt(\PBzs)>15\GeV$ (b).
The vertical band indicates the region $m_{\PX}\pm\Gamma_{\PX}$.
The lower panels display the pull (difference between the data and
the fit result, divided by the statistical uncertainty in the data).
\label{NominalFitBspi}}
\end{figure*}

A possible $\Xm$ signal contribution in the $ M^{\Delta}(\PBzs\Pgppm) $ spectrum
is modeled by a relativistic S-wave Breit--Wigner (BW) function, with mass
and width parameters fixed to $m_{\PX}$ and $\Gamma_{\PX}$, respectively. The
BW is convolved with a triple-Gaussian resolution function whose parameters
are obtained from the simulated data (standard deviation of the
triple-Gaussian function is about $2.2\MeV$ in the region of interest).
The background shape is approximated by a function of the form
$(x-x_{0})^\alpha \, \mathrm{Pol}_{n}(x)$,
where $x = M^{\Delta}(\PBzs\Pgppm)$, $x_{0} = m_{\PBzs} + m_{\Pgppm}$,
with $m_{\Pgppm}$ the $\Pgppm$ mass~\cite{PDG}, and $\mathrm{Pol}_{n}(x)$ represents a
polynomial function of order $n$. For the default shape $n=3$ is used.
The polynomial coefficients, as well as the
exponent $\alpha$ and the signal and background yields,
are obtained from the unbinned extended maximum-likelihood fit
shown in Fig.~\ref{NominalFitBspi}(a).
The fit returns a signal yield of $N_{\PX}= -85\pm160$ events.
The procedure is repeated requiring $\pt(\PBzs)>15\GeV$, and
the fit results displayed in Fig.~\ref{NominalFitBspi}(b)
give $N_{\PX}=-103\pm122$ events.%

Several cross-checks are performed and in all cases
the signal yield is consistent with zero. They include
repeating the fit with the following variations:
the background model parameters are fixed to the values
obtained from the fit with the \Xm signal region excluded;
the background model is fixed to the shape obtained from
simulated $\PBzs$ mesons combined with pion candidates
from the same simulated event;
different kinematic requirements and reconstruction
quality criteria are imposed on the
$\PBzs\Pgppm$, $\PBzs$, and $\Pgppm$ candidates;
collision events with multiple reconstructed
candidates are removed from the data sample,
and alternative background functions and fit regions
are used.

An upper limit on $\rho_{\PX}$, defined in Eq.~(\ref{eqnrho}),
is computed using the asymptotic
CL$_{S}$~\cite{Read:2002hq,JUNK1999435} method
developed in Ref.~\cite{CMS-NOTE-2011-005}.
The limit takes into account the following sources of
systematic uncertainty:
the uncertainty in the mass and the width of the
BW measured by the D0 Collaboration~\cite{D0Obs};
the uncertainty in $N(\PBzs)$;
the pion tracking efficiency uncertainty of $3.9\%$~\cite{TRK-11-001};
the uncertainty in $\epsilon_{\text{rel}}$ due to the finite number
of simulated events;
the description of the background by alternative
approximation functions, including the shape obtained
from simulation;
and modifications of the signal function due to
variations of the resolution function and the efficiency
with respect to $ M^{\Delta}(\PBzs\Pgppm) $ (both negligible).
The measured upper limit is
$\rho_{\PX} < 1.1\%$ at 95\% confidence level~(\CL) for the
baseline selection criteria ($\pt(\PBzs)>10\GeV$) and
$\rho_{\PX} < 1.0\%$ at 95\%~\CL for the
analysis requiring $\pt(\PBzs)>15\GeV$.
Using simulations of a spin-1 state decaying to $\PBstzs\Pgppm$,
where $\PBstzs\to\PBzs\gamma$ and where the mass is shifted by $m_{\PBstzs}-m_{\PBzs}$,
the upper limits were verified to differ negligibly between either
the spin-1 or spin-0 assumption.

\begin{figure}[h]
\begin{center}
\includegraphics[width=0.48\textwidth]{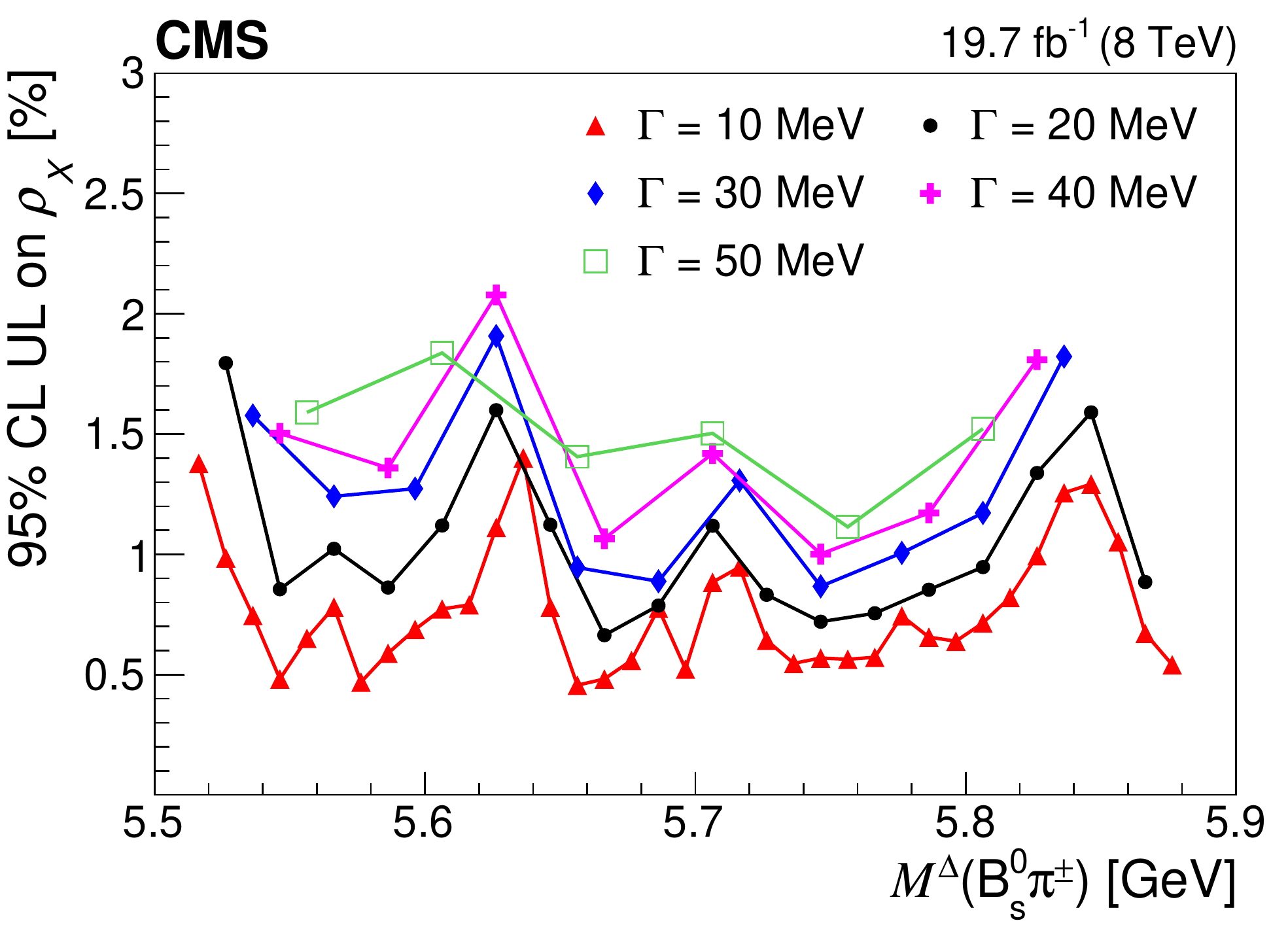}
\end{center}
\caption{
The $95\%$ \CL upper limit (UL) on $\rho_{\PX}$, Eq.~(\ref{eqnrho}), as a function
of the mass of a possible exotic state decaying into $ \PBzs\Pgppm $ for five different
values of the natural width of the state.
\label{fig:ULmass}}
\end{figure}

Upper limits are also obtained for different values of
mass and natural width ($\Gamma$) of a possible $\PBzs\Pgppm$ resonance,
as shown in Fig.~\ref{fig:ULmass}.
For these limits, no systematic uncertainties related to the mass and width
of the exotic state are assigned.
On the other hand, an additional systematic uncertainty in the relative
efficiency of up to 6\% is estimated for the extrapolation to high-mass
resonances from the low-mass simulation.
The limits are obtained for values of $\Gamma$ from $10$ to $50\MeV$
in $10\MeV$ steps,
while the mass takes values from $m_{\PBzs} {+} m_{\Pgppm} {+} \Gamma$
up to $5.9\GeV{-}1.5\Gamma$ in order to consider
a possible exotic state with higher
mass decaying to the $ \PBzs\Pgppm $ final
state~\cite{Esposito:2016itg,Ali:2016gdg}.
No significant excess is found throughout the region considered.

In summary, a search for the $\Xm$ state is performed by the CMS Collaboration
using $\Pp\Pp$ collision data collected at $\sqrt{s} = 8\TeV$ and corresponding
to an integrated luminosity of $19.7\fbinv$.
With about 50\,000 $\PBzs$ signal candidates,
no significant structure in the $\PBzs\Pgppm$
invariant mass spectrum is found around the mass reported
by the D0 Collaboration (nor for masses up to $5.9\GeV$).
The absence of a peak is supported by direct comparison with
the events in the $\PBzs$ sidebands, and by fits to the $\PBzs\Pgppm$
invariant mass distribution with
a resonant component included,
using different kinematic selection requirements,
as well as variants of the background modeling, fit regions,
and quality criteria.

Upper limits on the relative production rates
of the $\Xm$ and $\PBzs$ states, multiplied by the unknown branching
fraction of the $\Xmpm\to\PBzs\Pgppm$ decay, are computed to be:
\begin{linenomath*} \begin{align*}
&\rho_{\PX} <  1.1\%~\text{at}~95\%\text{ \CL  for }\pt(\PBzs)>10\GeV\text{ and}\\
&\rho_{\PX} <  1.0\%~\text{at}~95\%\text{ \CL  for }\pt(\PBzs)>15\GeV.
\end{align*} \end{linenomath*}

The upper limits on $\rho_{\PX}$ presented in this Letter are a factor of
two more stringent than the previous best limits, and do not
confirm the existence of the $\PX(5568)$ state.
These limits are also valid for a spin-1 state decaying into $\PBstzs\Pgppm$.
Additionally, upper limits
are set for different values of mass and natural width
of a hypothetical exotic resonance decaying into $\PBzs\Pgppm$.

We congratulate our colleagues in the CERN accelerator departments for the excellent performance of the LHC and thank the technical and administrative staffs at CERN and at other CMS institutes for their contributions to the success of the CMS effort. In addition, we gratefully acknowledge the computing centres and personnel of the Worldwide LHC Computing Grid for delivering so effectively the computing infrastructure essential to our analyses. Finally, we acknowledge the enduring support for the construction and operation of the LHC and the CMS detector provided by the following funding agencies: BMWFW and FWF (Austria); FNRS and FWO (Belgium); CNPq, CAPES, FAPERJ, and FAPESP (Brazil); MES (Bulgaria); CERN; CAS, MoST, and NSFC (China); COLCIENCIAS (Colombia); MSES and CSF (Croatia); RPF (Cyprus); SENESCYT (Ecuador); MoER, ERC IUT, and ERDF (Estonia); Academy of Finland, MEC, and HIP (Finland); CEA and CNRS/IN2P3 (France); BMBF, DFG, and HGF (Germany); GSRT (Greece); OTKA and NIH (Hungary); DAE and DST (India); IPM (Iran); SFI (Ireland); INFN (Italy); MSIP and NRF (Republic of Korea); LAS (Lithuania); MOE and UM (Malaysia); BUAP, CINVESTAV, CONACYT, LNS, SEP, and UASLP-FAI (Mexico); MBIE (New Zealand); PAEC (Pakistan); MSHE and NSC (Poland); FCT (Portugal); JINR (Dubna); MON, RosAtom, RAS, RFBR and RAEP (Russia); MESTD (Serbia); SEIDI, CPAN, PCTI and FEDER (Spain); Swiss Funding Agencies (Switzerland); MST (Taipei); ThEPCenter, IPST, STAR, and NSTDA (Thailand); TUBITAK and TAEK (Turkey); NASU and SFFR (Ukraine); STFC (United Kingdom); DOE and NSF (USA).

\hyphenation{Rachada-pisek} Individuals have received support from the Marie-Curie programme and the European Research Council and Horizon 2020 Grant, contract No. 675440 (European Union); the Leventis Foundation; the A. P. Sloan Foundation; the Alexander von Humboldt Foundation; the Belgian Federal Science Policy Office; the Fonds pour la Formation \`a la Recherche dans l'Industrie et dans l'Agriculture (FRIA-Belgium); the Agentschap voor Innovatie door Wetenschap en Technologie (IWT-Belgium); the Ministry of Education, Youth and Sports (MEYS) of the Czech Republic; the Council of Science and Industrial Research, India; the HOMING PLUS programme of the Foundation for Polish Science, cofinanced from European Union, Regional Development Fund, the Mobility Plus programme of the Ministry of Science and Higher Education, the National Science Center (Poland), contracts Harmonia 2014/14/M/ST2/00428, Opus 2014/13/B/ST2/02543, 2014/15/B/ST2/03998, and 2015/19/B/ST2/02861, Sonata-bis 2012/07/E/ST2/01406; the National Priorities Research Program by Qatar National Research Fund; the Russian Ministry of Education and Science, contracts 3.2989.2017 and 14.A12.31.0006; the Programa Severo Ochoa del Principado de Asturias; the Thalis and Aristeia programmes cofinanced by EU-ESF and the Greek NSRF; the Rachadapisek Sompot Fund for Postdoctoral Fellowship, Chulalongkorn University and the Chulalongkorn Academic into Its 2nd Century Project Advancement Project (Thailand); the Welch Foundation, contract C-1845; and the Weston Havens Foundation (USA).

\bibliography{auto_generated}
\ifthenelse{\boolean{cms@external}}{}{
\clearpage
\appendix
\section{Effect of \texorpdfstring{$\Delta R(\PBzs,\Pgppm)$}{delta R(B0s, pi+-)}
requirement on the \texorpdfstring{$M^{\Delta}(\PBzs\Pgppm)$}{M(B0s pi+-)} distribution\label{app:suppMat}}

The D0 Collaboration obtains a significance above five
standard deviations for the $\Xm$ state only when an upper limit on
$\Delta R=\sqrt{\smash[b]{(\Delta\eta)^2 + (\Delta\phi)^2}}$
is imposed, namely, $\Delta R < 0.3$.
Here, $\Delta\eta$ and $\Delta\phi$ are the pseudorapidity and azimuthal
angle (in radians) differences between the directions of the $\PBzs$ and $\Pgppm$.
As shown in Fig.~\ref{BspiSigDeltaR}, a requirement on this variable
significantly changes the $M^{\Delta}(\PBzs\Pgppm)$ distribution
and can even produce a peaking shape;
therefore, it is not used in this analysis.
\begin{figure}[hpb]
\begin{center}
\includegraphics[width=0.48\textwidth]{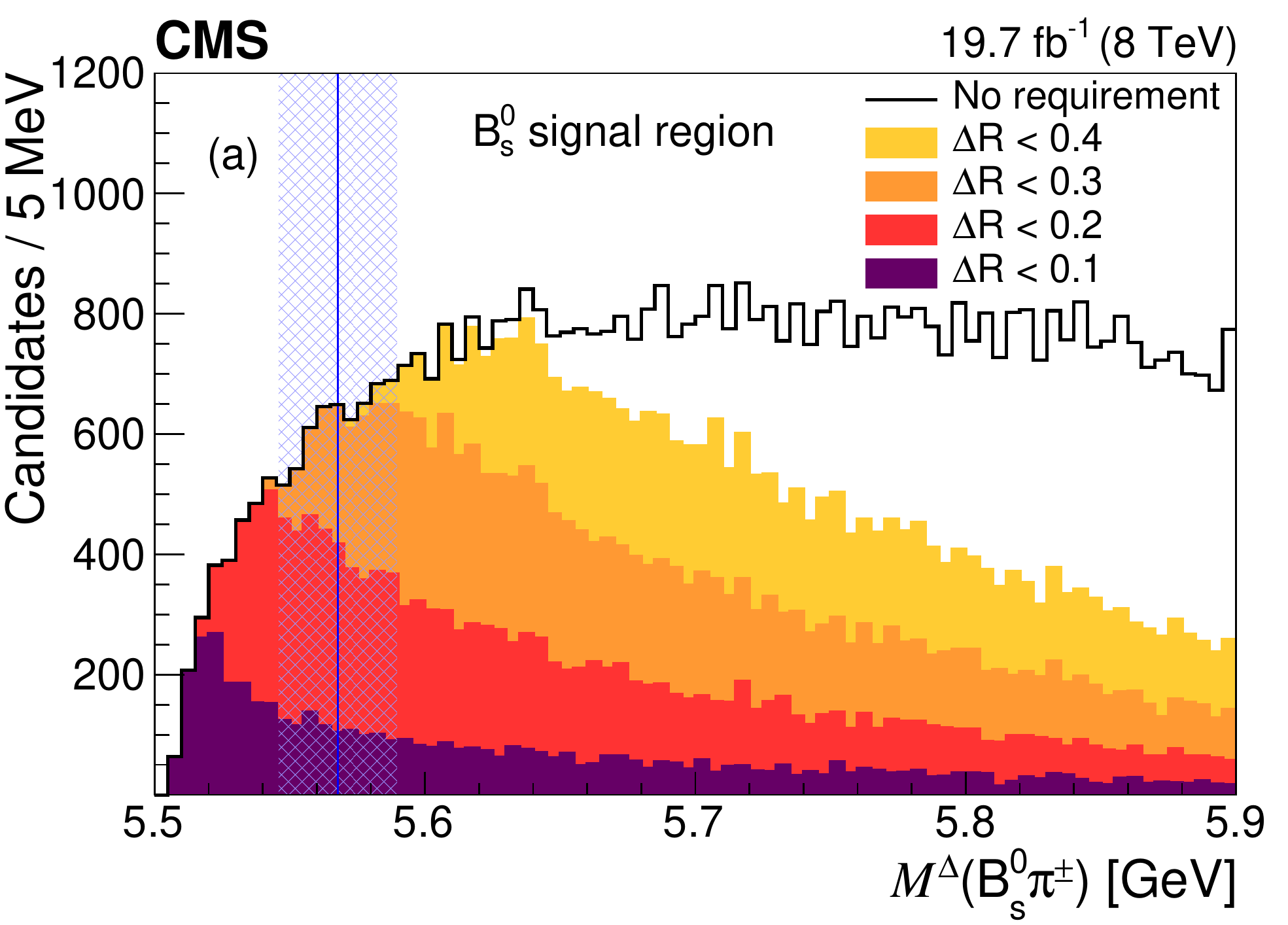}
\includegraphics[width=0.48\textwidth]{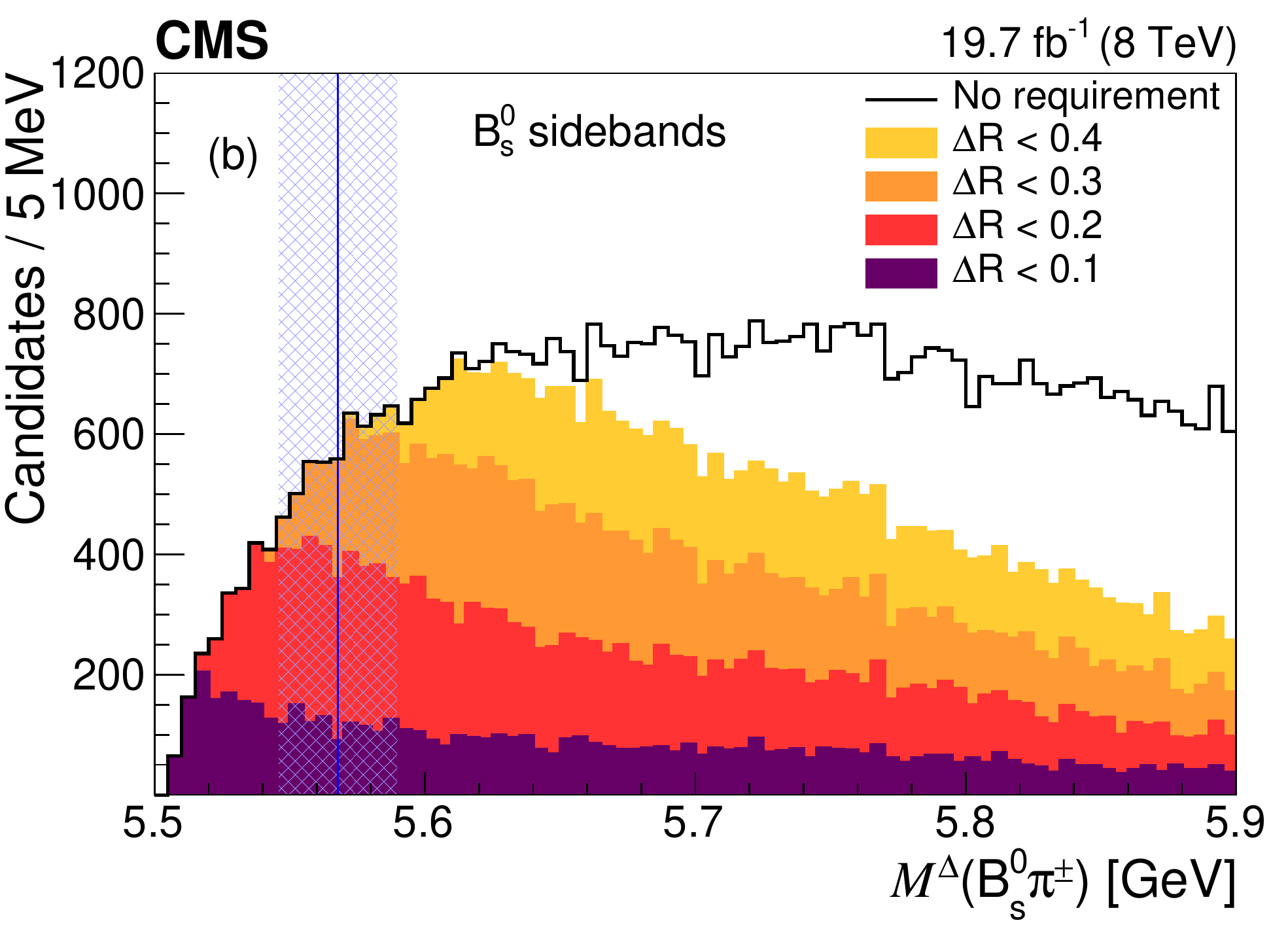}
\end{center}
\caption{
The $M^{\Delta}(\PBzs\Pgppm)$ distributions for events in the $\PBzs$ (a) signal
and (b) sideband regions for different
$\Delta R$ requirements.
The uncertainties are not shown for the sake of clarity.
The vertical band indicates the region $m_{\PX}\pm\Gamma_{\PX}$.
\label{BspiSigDeltaR} }
\end{figure}
}
\cleardoublepage \section{The CMS Collaboration \label{app:collab}}\begin{sloppypar}\hyphenpenalty=5000\widowpenalty=500\clubpenalty=5000\input{BPH-16-002-authorlist.tex}\end{sloppypar}
\end{document}

%% file: BPH-16-002-authorlist.tex
\textbf{Yerevan Physics Institute,  Yerevan,  Armenia}\\*[0pt]
A.M.~Sirunyan, A.~Tumasyan
\vskip\cmsinstskip
\textbf{Institut f\"{u}r Hochenergiephysik,  Wien,  Austria}\\*[0pt]
W.~Adam, F.~Ambrogi, E.~Asilar, T.~Bergauer, J.~Brandstetter, E.~Brondolin, M.~Dragicevic, J.~Er\"{o}, A.~Escalante Del Valle, M.~Flechl, M.~Friedl, R.~Fr\"{u}hwirth\cmsAuthorMark{1}, V.M.~Ghete, J.~Grossmann, J.~Hrubec, M.~Jeitler\cmsAuthorMark{1}, A.~K\"{o}nig, N.~Krammer, I.~Kr\"{a}tschmer, D.~Liko, T.~Madlener, I.~Mikulec, E.~Pree, N.~Rad, H.~Rohringer, J.~Schieck\cmsAuthorMark{1}, R.~Sch\"{o}fbeck, M.~Spanring, D.~Spitzbart, W.~Waltenberger, J.~Wittmann, C.-E.~Wulz\cmsAuthorMark{1}, M.~Zarucki
% ...
\vskip\cmsinstskip
% ...
1:~~Also at Vienna University of Technology, Vienna, Austria\\
% ...
